# THE INFLUENCE OF VERTICAL RESOLUTION ON INTERNAL TIDE ENERGETICS AND SUBSEQUENT EFFECTS ON UNDERWATER ACOUSTIC PROPAGATION

Luna Hiron[1], Martha C. Schönau[2], Keshav J. Raja[1], Eric P. Chassignet[1], Maarten. C. Buijsman[3], Brian K. Arbic[4], Alex Bozec[1], Emanuel M. C. Coelho[5], and Miguel Solano[6]

[1]Center for Ocean-Atmospheric Prediction Studies, Florida State University

[2]Scripps Institution of Oceanography, University of California, San Diego

[3]School of Ocean Science and Engineering, The University of Southern Mississippi

[4]Department of Earth and Environmental Sciences, University of Michigan

[5]Applied Ocean Sciences (AOS), LLC

[6]SOFAR Ocean Technologies

Corresponding author: Luna Hiron (lhiron@fsu.edu)

## ABSTRACT

Internal tide generation and breaking play a primary role in the vertical transport and mixing of heat and other properties in the ocean interior, thereby influencing climate regulation. Additionally, internal tides increase sound speed variability in the ocean, consequently impacting underwater acoustic propagation. With advancements in large-scale ocean modeling capabilities, it is essential to assess the impact of higher model resolutions (horizontal and vertical) in representing internal tides. This study investigates the influence of vertical resolution on internal tide energetics and its subsequent effects on underwater acoustic propagation in the HYbrid Coordinate Ocean Model (HYCOM). An idealized configuration with a ridge, forced only by semidiurnal tides and having 1-km horizontal grid-spacing, is used to test two different vertical-grid discretizations, defined based on the zero-crossings of horizontal velocity eigenfunctions, with seven distinct numbers of isopycnal layers, ranging from 8 to 128. Analyses reveal that increasing the number of layers up to 48 increases barotropic-to-baroclinic tidal conversion, available potential energy, and vertical kinetic energy, reaching equilibrium afterwards with higher layer counts. Vertical shear exhibits a similar pattern but converging at 96 layers. Simulations with at least 48 layers fully resolve the available potential energy contained in the $3^{rd}$ to $8^{th}$ tidal baroclinic modes. Finally, sound speed variability and acoustic parameters differ for simulations with less than 48 layers. Therefore, the study concludes that a minimum vertical resolution (48 layers in this case) is required in isopycnal models to minimize the impact on internal tide properties and associated underwater acoustic propagation.

# 1. Introduction

Internal tide generation and breaking play a primary role in the vertical transport and mixing in the ocean interior (Munk and Wunsch, 1998; Ferrari and Wunsch, 2009; Vic et al., 2019). The vertical isopycnal displacement induced by internal-tidal-induced can be as large as ~200 *m*, with the highest displacements usually observed in the thermocline (Park et al., 2008; Klymak et al., 2011; Rainville et al., 2013). The breaking of internal tides drives significant diapycnal mixing and vertical heat and salt fluxes, resulting in substantial impacts on the heat content of the upper ocean, thereby influencing climate regulation (Hebert, 1994; Storlazzi et al., 2020). This high-frequency variability induced by internal tides holds significant importance for mixing parameterizations in global climate models. Another significant outcome is the vertical movement of nutrients, essential for primary production and increased carbon uptake, which in turn has further implications for climate dynamics (Tuerena et al., 2019; Kossack et al., 2023). Moreover, internal tides increase sound speed variability in the ocean, which, in turn, affects underwater acoustic propagation, as reviewed recently in Schonau et al. (2024), and also in Yang et al. (2010), Colosi et al., (2013), Turgut et al. (2013), Noufal et al. (2022), among others. Changes in the underwater acoustic propagation have direct applications for sonar performance precision, bioacoustic source localization, acoustic tomography, and underwater acoustic communication.

Internal (baroclinic) tides are internal waves generated by the interaction of barotropic tides with topography features in the stratified ocean environment (Wunsch, 1975; St. Laurent and Garrett, 2002). These waves can radiate very far (> 1000 *km*) from their region of origin (Dushaw et al., 1995; Ray and Mitchum, 1996, 1997; Rainville and Pinkel, 2006; Alford and Zhao, 2007; Buijsman et al., 2016, 2020), and up to 50% of the total baroclinic tidal energy can dissipate locally through wave breaking, wave-wave interactions, and scattering towards higher harmonics and wavenumbers (St. Laurent and Garrett, 2002; Lamb, 2004; Vic et al., 2019; Eden et al., 2020; Solano et al., 2023). Increase in dissipation also occurs when internal tides interact with mesoscale eddies, shifting these waves from stationary to non-stationary (e.g., Ray and Zaron, 2011; Zaron and Egbert, 2014; Shriver, et al., 2014; Ponte and Klein, 2015; Buijsman et al., 2017; Zaron, 2017; Nelson et al., 2019; Wang and Legg, 2023; Yadidya et al., 2024; Delpeche et al., 2024).

Recent computational advancements have allowed the inclusion of tides in high-resolution global ocean models (Arbic et al., 2012, 2018; Arbic, 2022). However, the variation in model parameters and grid-spacing can affect how internal tides are represented in these models and their consequent wave-wave and wave-mean flow interactions. For example, the increase in bathymetry resolution generates stronger internal tides at a local scale (Xu et al., 2023). Furthermore, Buijsman et al. (2020) showed that decreasing the horizontal grid spacing from 8 km to 4 km in realistic, global HYbrid Coordinate Ocean Model (HYCOM) simulations increased the semidiurnal barotropic-to-baroclinic tidal conversion by 50%. This enhancement in tidal conversion is associated with an increase in the number of vertical modes resolved (from 1-2 to 1-5 $M_2$ modes) and a better representation of wave-wave interactions.

Nelson et al. (2020) found that decreasing the horizontal grid spacing in regional Massachusetts Institute of Technology general circulation model (MITgcm) simulations from 2 to 0.25 *km* greatly improves the internal wave frequency spectra, and that decreasing only the vertical grid spacing by a factor of three (from 90 to 270 depth levels) does not yield any significant improvement. However, increasing both vertical and horizontal grid spacing yielded the best comparison of internal wave frequency spectra to observations. Also using regional MITgcm simulations, Thakur et al. (2022) found that increasing the vertical grid spacing from 109 to 264 depth levels for a horizontal grid spacing of 1/48º (~ 2 *km* in their domain) improved the representation of small-vertical-scale density and velocity fluctuations, improving the internal wave (IW) field, which, in turn, better represented IW-induced mixing and dispensed the need for the background value of the KPP mixing-parameterization.

Although the impact of increasing vertical spacing grid on internal tides in MITgcm has been investigated, an in-depth study on the impact of the increase in vertical grid-spacing in HYCOM, which uses isopycnal coordinates in the stratified ocean interior as opposed to constant depth coordinates (or z-levels) as in MITgcm, is still lacking. Thanks to the isopycnal coordinate system, HYCOM needs significantly fewer vertical layers compared to z-level models to resolve the same number of vertical modes (Buijsman et al., 2020, 2024; Xu et al., 2023). Using several criteria to determine how many modes are resolved, Buijsman et al. (2024) predicts that a realistically forced global HYCOM simulation with 41 hybrid vertical coordinates and 1/25º (~4 *km*) horizontal grid-spacing resolves on average 6 to 12 modes, with the number of modes resolved decreasing poleward due to the reduction in stratification and the increase in the layer thickness in the ocean interior. They also found that for this global HYCOM simulation the limiting factor for the resolution of vertical modes for the dominant semidiurnal ($M_2$) internal tides in the tropics is the horizontal grid-spacing, whereas at higher latitudes the vertical grid-spacing becomes the limiting factor. With the continuous surge in computational power, it has become imperative to understand the implications of increasing resolutions and the benefits they bring to simulations, in an attempt to find the optimal equilibrium between computational efficiency and efficacy.

There has been an increase in the number of submesoscale-resolving regional and basin-scale HYCOM simulations developed in the past years, such as the 1/100° (~1 *km*) horizontal grid-spacing simulation for Gulf of Mexico and 1/50° (~2 *km*) simulation for the North Atlantic (Chassignet and Xu, 2017; Hiron et al., 2021, 2022; Uchida et al. 2022; Xu et al., 2023; Chassignet et al., 2023). Recent discussions among oceanographers and ocean modelers center on performing a global HYCOM simulation with tidal forcing and finer horizontal grid-spacing on the order of 1/50°, and even potentially 1/100° in the future, to replace the current state-of-the-art global HYCOM with 1/25º grid spacing. However, the optimal number of layers for such simulation is still an open question, with a debate surrounding the number of layers needed to resolve a given number of baroclinic modes. Xu et al. (2023) shows that, in theory, a HYCOM simulation with sufficient horizontal resolution only requires two layers (i.e., one interface depth) to resolve the first baroclinic Rossby radius of deformation, as long as the interface is placed at the depth of the zero-crossing of the 1st baroclinic mode of the horizontal velocity eigenfunction. Similarly, only two interface depths are needed to resolve the second baroclinic Rossby radius of deformation, and so on. Thus, according to Xu et al. (2023), only *n* number(s) of interface depths are needed to resolve the nth baroclinic mode of meso- and large scale motions. However, they also find that three interface depths (3*n*) are required for the maximum amount of energy (kinetic and available potential energy) permitted by a given horizontal grid spacing to be projected onto the first baroclinic mode, assuming that neither horizontal nor vertical grid-spacing are limiting. In other words, one interface depth will "allow" energy to be projected onto the first baroclinic mode, but three interface depths are needed to maximize the energy in this mode provided that horizontal grid-spacing and number of layers are not limiting.

For internal tides, Buijsman et al. (2024) test different criteria to determine when a vertical mode is resolved by the vertical number of layers. They argue that for the global 1/25° grid-spacing HYCOM simulation, a given mode cannot be resolved if two horizontal velocity eigenfunction (*u*- eigenfunctions) zero-crossings occur within the same isopycnal layer. Thus, according to Buijsman et al. (2024), at least n interface-depths is required to resolve the nth baroclinic mode, provided that only one $u$-eigenfunction zero-crossing is located in each layer. This criteria is less strict than Xu et al. (2023) to some extent, as the interface depths were not required to coincide with the zero-crossing of a given $u$-eigenfunction. Once again, by "resolving" we mean energy being projected onto the modes, but not necessarily the maximum amount permitted by the horizontal grid-spacing. For this manuscript, we want to test the optimal number of layers that not only "allow" energy to be projected onto the maximum number of baroclinic modes, but that also maximize the energy in this mode permitted if both horizontal grid-spacing and number of layers are not limiting.

This paper investigates the impact of the increase in the number of vertical layers in a ~1-*km* grid-spacing HYCOM simulation on (1) the representation of internal tide energetics (kinetic energy and available potential energy) and squared vertical shear, (2) the number of tidal baroclinic modes resolved and the amount of kinetic energy and available potential energy contained in the lower and higher modes, and (3) the subsequent effects on sound speed variability and underwater acoustic propagation. The above will be investigated using an idealized configuration of HYCOM with a 1/100º horizontal grid-spacing, forced solely with semidiurnal barotropic tide and having a density profile characteristic of the tropics. Different numbers of vertical layers (all isopycnal) and two different types of grid discretization will be employed. The idealized configuration allows us to isolate the internal wave problem and avoid "contamination" from meso- and large scale motions, wind-driven near-inertial waves, and wave-mean flow interactions. One important characteristic that we are not investigating in this study is the internal wave continuum frequency and vertical wavenumber spectrum, which is also impacted by vertical resolution (e.g., Nelson et al. 2020, Thakur et al. 2022), but which only arises in a configuration with extensive wave-wave interactions. The impact of vertical resolution on continuum spectra and mixing, that are sensitive to very short scales, may differ from the impact of vertical resolution on bulk quantities considered in internal tide energetics, which are contained in the lower tidal vertical modes.

## 2. Methodology

### 2.1. HYbrid Coordinate Ocean Model (HYCOM)

The HYbrid Coordinate Ocean Model (HYCOM) is a hydrostatic ocean general circulation model system (Bleck, 2002). Recent modeling advancements, such as smaller grid-spacing and hourly outputs, have allowed the inclusion of both barotropic and internal tidal components into HYCOM and have expanded our understanding of tidal dynamics on a global scale (Arbic et al., 2012, 2018; Arbic, 2022). HYCOM uses isopycnal coordinates in the stratified ocean interior, pressure coordinates near the surface and in the mixed layer, and terrain-following coordinates on the shelves (Chassignet et al., 2003; 2009). HYCOM's unique vertical coordinate system distinguishes it from other global models that use z-level coordinates such as MITgcm or the European model NEMO. In our idealized simulations, only isopycnal coordinates were used.

### 2.2. Idealized configuration

An idealized configuration is chosen to study the effects of vertical resolution on internal tide energetics without potential contamination from meso- and large-scale motions. The idealized configuration consists of a two-dimensional box, with 1/100º horizontal grid-spacing, 8000 grid-points in the longitudinal direction, and 4000 *m* depth. The simulations use HYCOM version 2.3.01, has hourly outputs, and is only forced by the semidiurnal ($M_2$) tidal constituent. M2 is the largest lunar constituent and its internal tide field contains about 70% of all tidal energy (Egbert and Ray, 2003). The amplitude of the barotropic tide used here is equal to the tidal amplitude of the Amazon shelf, an area known for its relatively large tides. The simulations are initialized with a generic density profile representative of the tropic. To avoid freshwater influence from the Amazon shelf region, the "generic" density profile from the tropics is obtained by averaging World Ocean Atlas 2018 climatology (Garcia et al., 2019) over 17° N – 23° N and 27° W – 12° W during the summer months (Figure 1a,b), referenced to the surface (sigma zero; $\sigma_0$). This density profile is used for the initial and boundary conditions. A ridge with a Gaussian shape was added in the center of the domain with a height of 3500 *m* and a standard deviation (also called Gaussian root mean square width) of 17 *km* (Figure 1c).

Some non-dimensional parameters are provided below to characterize the regime of the baroclinic tides, following Garrett and Kunze (2007) and Buijsman et al. (2010). These parameters are computed based on the following variables of our configuration: the amplitude of the tidal barotropic velocity $U_0$ over the ridge is ~0.05 $m\,s^{-1}$, the maximum ridge height $H$ is 3500 *m*, the maximum buoyancy frequency $N_{max}$ is 0.015 $s^{-1}$, and the topographic length scale (Gaussian standard deviation; $\sigma$ or $L$) is 17 *km*. The criticality of the slope $\gamma$ is expressed as $\gamma = \max\left(\frac{1}{\alpha}\frac{\partial h}{\partial x}\right)$, where $h(x)$ is the topographic height, and $\alpha = \sqrt{\frac{\omega^2 - f^2}{N^2 - \omega^2}}$, with $\omega$ being the M2 tidal frequency, $f$ the Coriolis frequency (near zero for our case), and $N$ the buoyancy frequency. The slope parameter for our configuration is ~2.3, i.e., supercritical ($\gamma > 1$), which means the ridge slope is steep enough to generate nonlinear waves and well-defined internal wave beams that are directed diagonally downward (Balmforth et al., 2002; Garrett and Kunze, 2007; Buijsman et al., 2010). Strong velocity shear along the beams is also presented with this regime. The tidal excursion number [$Ex = U_0/(L\omega)$] associated with our configuration is small, ~0.02 ($Ex \ll 1$), which guarantees the generation of coherent internal wave beams, with baroclinic velocity substantially larger than barotropic tidal velocity and wave beam angle associated with the M2-tidal barotropic forcing frequency (Jalali et al., 2014). The topographic Froude number [$Fr_t = U_0/(N_{max}H)$] is also small, ~1x10$^{-4}$ ($Fr_t \ll 1$), which means that the flow is affected by the topography and blocking occurs. Thus, with $\gamma > 1$, $Ex \ll 1$, and $Fr_t \ll 1$, our configuration falls between regimes 4 and 5 of Garrett and Kunze (2007). This regime includes nonlinear internal hydraulic jumps and generation of internal waves at higher harmonics of the forcing frequency, and is similar to the regime found in the Luzon Strait in Buijsman et al. (2010).

To document the effect of the vertical resolution on internal tides, multiple simulations are carried out keeping all other parameters constant and only varying the numbers of isopycnal layers and grid discretization. Two sets of vertical grid discretization are tested to examine the dependency of our results on the way the layers are distributed. In the first set of experiments, the interface depths are defined using the zero-crossings of the horizontal velocity eigenfunctions (also called *u*-eigenfunctions; Kelly, 2016; Kelly and Lermusiaux, 2016) based on the density profile (Figure 1a). For a simulation of n layers, the interface depths are defined as the depths of the zero-crossings of the (*n-1*)*th* mode *u*-eigenfunction for the ocean interior, plus the "surface", considered as the interface at pressure equals zero, and the bottom, providing then a vertical grid with n number of layers. The logic behind the usage of *u*-eigenfunctions to define the layers come from the fact that to resolve a specific mode number in an isopycnal model, the interface layers should coincide with the zero-crossings of the *u*-eigenfunction, ensuring that the maximum horizontal velocity occurs within each layer (Buijsman et al., 2024; Xu et al., 2023). Although we are not solving more than the first few baroclinic modes in these simulations due to limitations in horizontal and vertical resolution (Buijsman et al., 2020), this technique is an objective way to define the interface depths of our simulations. Seven simulations are performed with the following number of vertical layers: 8, 16, 32, 48, 64, 96, 128 (Figure 2). For the second set of simulations, we start with the 128-layer simulation described above, and merge consecutive layers to obtain vertical grids with the following number of layers: 8, 16, 32, and 64 layers (Figure 3). See Section 2.3 for more information on the vertical mode decomposition.

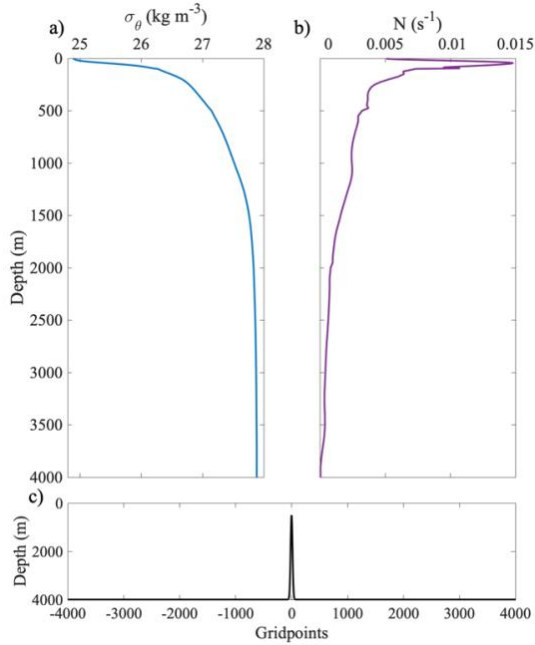

Figure 1. a) Density profile used as the initial condition for all simulations and associated b) buoyancy frequency. c) Topography of the idealized configuration, with a 3500 m high ridge in the middle of the domain.

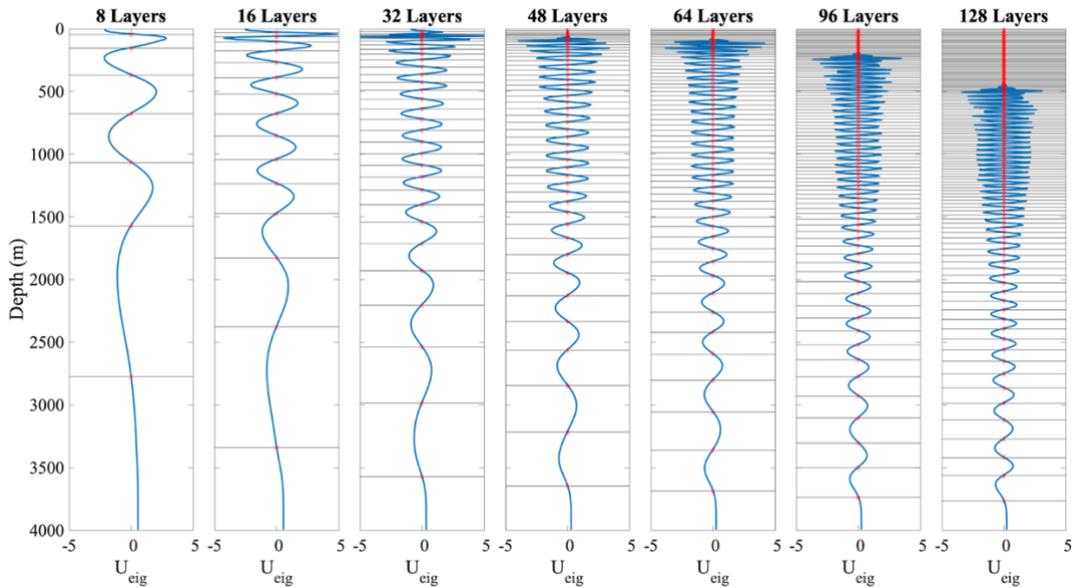

Figure 2. Horizontal velocity eigenfunctions (blue curves) with the location of the zero-crossings (red stars). The black horizontal lines intersecting the red stars are the interface depths used to initialize the first set of simulations with 8, 16, 32, 48, 64, 96, and 128 layers.

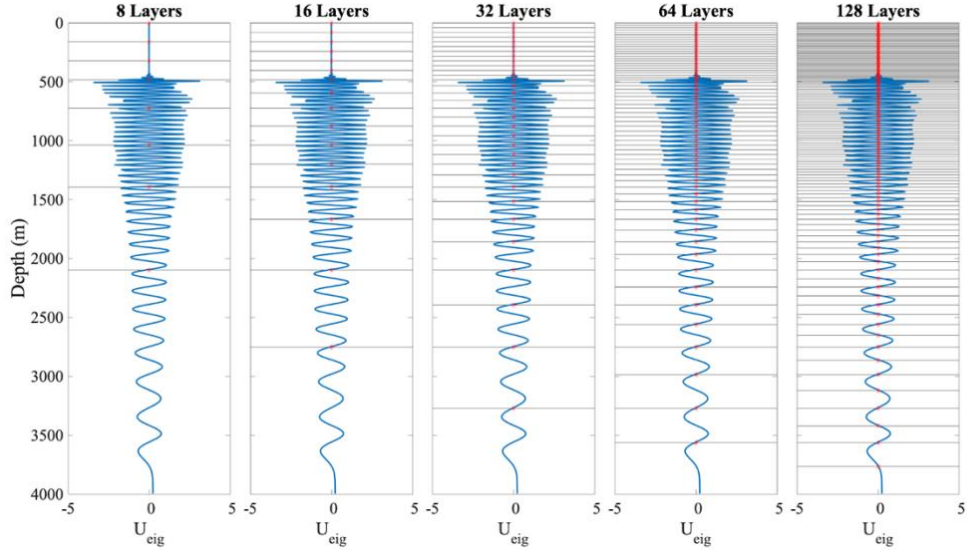

Figure 3. Distribution of interface-depths (black horizontal lines) for the second set of simulations based on merging consecutive layers starting from the 128-layer simulation defined by the zero-crossings of the *u*-eigenfunction (blue curves in all subplots) with 8, 16, 32, 64, and 128 layers. Please note that the 128-layer simulation is the same in both sets.

Zonal boundaries feature a relaxation time ranging from 0.1 to 1 day, accompanied by high viscosity, to avoid reflection on the zonal boundaries. All vertical physics, including the K-Profile Parameterization (KPP) scheme, are turned off. Note that the publicly available version of HYCOM does not support strict 2-dimensional simulations. To address this, we introduced five additional grid-points in the meridional direction and enforced a zero meridional velocity (*v*) at each time step. This approach allows us to maintain a computationally efficient simulation while preventing wave reflections in the meridional direction. The configuration is symmetric with respect to the ridge; thus, for practicality, the energy diagnostics are done on one side of the domain and integrated from the ridge to a point 250 *km* from the ridge. The simulations are run for 30 days with hourly outputs (total of 721 time steps). All analyses, except for the KE spectra, are conducted over four tidal periods after the model reaches equilibrium, which occurs after 8 days. The KE spectra is computed over the remaining days after equilibrium, a total of 22 days.

### 2.3. Vertical mode decomposition

Baroclinic fields can be decomposed into vertical standing waves (normal modes) that propagate horizontally (Gerkema and Zimmerman, 2008; Kelly et al., 2012; Kelly, 2016; Buijsman et al., 2014; Buijsman et al., 2020, Early et al., 2021; Raja et al., 2022). The vertical velocity eigenfunction $\mathcal{W}_n(z)$ of the mode *n* can be found by solving the following Stürm-Liouville equation

$$\frac{\partial^2 \mathcal{W}_n(z)}{\partial z^2} + \frac{N^2}{c_n^2} \mathcal{W}_n(z) = 0 \qquad (1)$$

where $c_n$ is the eigenspeed, $N = \sqrt{-\frac{g}{\rho_0}\frac{\partial \rho(z)}{\partial z}}$ is the buoyancy frequency computed using the initial potential density profile $\rho$ referenced to the surface (Figure 1a), $g$ is the acceleration of gravity, $\rho_0$ is the constant density associated with the Boussinesq approximation, taken here as 1027 kg m$^{-3}$, and z is the vertical coordinate. The eigenspeed is expressed as

$$c_n = \frac{\sqrt{\omega^2 + f^2}}{k_n} \qquad (2)$$

where $\omega$ and $f$ are the M$_2$ and Coriolis frequencies, respectively, and $k_n$ is the horizontal wavenumber. The horizontal velocity eigenfunction (or *u*-eigenfunction) is then found by taking the derivative of $\mathcal{W}_n(z)$ in the *z* direction,

$$\mathcal{U}_n(z) = \frac{\partial \mathcal{W}_n}{\partial z}(z), \qquad (3)$$

and normalizing it by the depth averaged $\mathcal{U}_n$ amplitude, i.e., $\sqrt{\frac{1}{H}\int_0^H \mathcal{U}_n^2(z)dz}$, where $H$ is the water column depth, following the same method as previous studies (Gill, 1982; Gerkema and Zimmerman, 2008; Kelly et al., 2012; Kelly, 2016; Buijsman et al., 2014; Buijsman et al., 2020, Early et al., 2021; Raja et al., 2022). The *u*-eigenfunction for mode numbers 8, 16, 32, 48, 64, 96, and 128 (blue lines) and the corresponding zero-crossings (horizontal black lines) are shown in Figure 2.

### 2.4. M$_2$ energetics and vertical shear

All quantities described below were computed in the original HYCOM grid (isopycnal layers), including tidal energetics, squared vertical shear, barotropic-to-baroclinic conversion, and dissipation. The only quantities computed on interpolated fields are the *u*- and *w*- eigenfunctions and modal kinetic energy and available potential energy (section 2.4.4). These were estimated by interpolating the density profile in the original vertical grid into an equidistant vertical grid, with a one meter vertical grid-spacing, using a Piecewise Cubic Hermite Interpolating Polynomial.

*2.4.1. Tidal barotropic-to-baroclinic energy conversion and dissipation*

Tidal barotropic-to-baroclinic energy conversion is computed as in Kelly et al. (2012):

$$C(x,t) = p'_{bottom}(x,t)\, u_{btp}(x,t)\, \frac{\partial H(x)}{\partial x} \qquad (4)$$

where $u_{btp}$ is the barotropic tidal velocity, $p'_{bottom}$ is the perturbation pressure at the bottom, and $H$ is the depth of the water column. The perturbation pressure is computed by removing the time-mean and the depth-mean pressure from the pressure field. The vertically-integrated baroclinic energy flux $F_p$ can be computed as the product of the perturbation baroclinic velocity $u'_{bcl}$ and the perturbation pressure $p'$ integrated vertically in the water column:

$$F_p(x,t) = \int_0^z u'_{bcl}(x,z,t)\, p'(x,z,t)\, dz. \qquad (5)$$

The perturbation baroclinic velocity is computed by removing the time-mean velocity from the velocity field. The residual between the tidal barotropic-to-baroclinic energy conversion and the baroclinic energy flux divergence integrated over a domain provides an indirect estimation of the amount of energy dissipated D locally/within the domain:

$$D = C - \nabla \cdot F_p \qquad (6)$$

### 2.4.2. Kinetic energy and available potential energy

The available potential energy ($APE$) is computed following Gill (1982) and Kundu (1990):

$$APE = \frac{1}{2}\rho_0 N^2 \zeta^2 \tag{7}$$

where $N$ is the buoyancy frequency at time zero, and $\zeta$ is the displacement of the isopycnals relative to its position at time zero. Kang and Fringer (2010) remarked that the equation above ($APE_3$ in their paper), derived from linear theory, is used in internal wave calculations for slowly varying density fields, and that another equation ($APE_2$ in their paper) is more suitable to account for strong nonlinear and nonhydrostatic effects. In this paper, the results using the two equations were almost identical, in agreement with the fact that HYCOM is a hydrostatic model and that our density profile does not present any "too" sharp vertical density gradient (the domain has a horizontally uniform stratification). Thus we choose to use the APE equation presented above following Gill (1982) and Kundu (1990).

The total baroclinic kinetic energy $KE$ is computed as follows:

$$KE_{bcl} = \frac{1}{2}(u_{bcl}^2 + w^2) \tag{8}$$

where $u_{bcl}$ is the baroclinic zonal velocity, and $w$ is the vertical velocity. The vertical $KE$ is computed using only the vertical velocity $w$.

### 2.4.3. Squared vertical shear

Squared vertical shear $S^2$ is computed using both the baroclinic zonal and vertical velocity components, as follows

$$S^2 = \left(\frac{\partial u_{bcl}}{\partial z}\right)^2 + \left(\frac{\partial w}{\partial z}\right)^2. \tag{9}$$

Conversion, baroclinic $KE$, $APE$, and vertical shear were computed after the model reached a stable state, which was reached after 200 hours.

### 2.4.4. Modal energetics

Following Gerkema and Zimmerman (2008), Kelly et al. (2012), and Buijsman et al. (2014, 2020), and using equations 1 and 3, we can find how much $KE$ and $APE$ is present in each vertical mode. For that, first, we compute the modal amplitudes of the zonal velocity by projecting the $u$-eigenfunctions onto the vertical profiles of horizontal velocity:

$$\hat{u}_n = \frac{1}{H}\int_0^H \mathcal{U}_n(z)\, u(z)\, dz. \tag{10}$$

As done in Raja et al. (2022), we multiply the $u$-eigenfunction by the zonal velocity $u$. We then compute the horizontal velocity associated with each mode by multiplying the modal amplitudes by the $u$-eigenfunctions:

$$u_n(z) = \hat{u}_n \mathcal{U}_n(z). \tag{11}$$

The modal amplitude of the vertical velocity associated with mode $n$ is as follows:

$$\widehat{w}_n = \frac{1}{H} \int_0^H \mathcal{W}_n(z)\, w(z)\, N^2(z)\, dz, \tag{12}$$

$$w_n(z) = \widehat{w}_n \mathcal{W}_n(z). \tag{13}$$

The isopycnal vertical displacement $\zeta$ is:

$$\widehat{\zeta}_n = \frac{1}{H} \int_0^H \mathcal{W}_n(z)\, \zeta(z)\, N^2(z)\, dz, \tag{14}$$

$$\zeta_n(z) = \widehat{\zeta}_n \mathcal{W}_n(z). \tag{15}$$

Modal $KE$ and $APE$ are then computed following equations 7 and 8.

**2.5. Sound speed and underwater acoustic propagation**

*2.5.1. Sound speed*

The sound speed $ssp$ ($m\ s^{-1}$) was computed following Mackenzie (1981)'s equation:

$$ssp = 1448.96 + 4.591\, T - 5.304 \times 10^{-2}\, T^2 + 2.374 \times 10^{-4}\, T^3 +$$
$$(S - 35)(1.340 - 1.025 \times 10^{-2}\, T) + 1.630 \times 10^{-2}\, Z +$$
$$1.675 \times 10^{-7}\, Z^2 - 7.139 \times 10^{-13}\, T\, Z^3.$$

where $T$ is the in situ temperature in °C, converted from the HYCOM potential temperature, $S$ is salinity, and $Z$ is depth (positive values). Temperature and salinity were first interpolated to 1 m depth surfaces, using a Piecewise Cubic Hermite Interpolating Polynomial, prior to calculating sound speed.

*2.5.2. Underwater acoustic propagation*

Using the model sound speed, Bellhop 3D was used to model acoustic propagation. Bellhop 3D is available from the Ocean Acoustics Laboratory Acoustic Toolbox (http://oalib.hlsresearch.com/AcousticsToolbox; Porter, 2011). Bellhop is a ray-tracing model that can trace propagation pathways using either 3D or 2D pressure fields. Bellhop 3D was run for a 1500 Hz source placed at 20 m depth at the ridge using each of the 18 HYCOM model hourly time steps. The model was run in semi-coherent mode, to increase sensitivity to ray phases, and output acoustic transmission loss (TL), a measure of acoustic loss from both attenuation and spreading (Urick, 1982). The idealized HYCOM model output for each time stamp, which is strictly zonal, was replicated in the meridional direction so that some 3D acoustic impacts may be seen. The goal was to examine the sensitivity of the upper-ocean sound speed structure and acoustic propagation to the internal tide layers at relatively short ranges (<150 *km*). Additionally, we calculated the sonic layer depth (SLD), the depth of subsurface sound speed maximum above which an acoustic duct can form, the below-layer gradient (BLG), the gradient in sound speed in th the 100 m transitional layer below the SLD, and the in-layer gradient (ILG), defined as the gradient of sound speed in the sonic layer. These can at times be indicators for surface-layer duct propagation (Urick, 1982; Helber et al., 2012; Colosi and Rudnick 2020).

## 3. Tidal energetics and squared vertical shear

### 3.1. Baroclinic kinetic energy and available potential energy spatial patterns

A snapshot of the baroclinic velocity field for each of the 8- and 128-layer simulations, with grid-spacing defined based on the *u*-eigenfunctions, are shown in Figure 4a,b. The wave beams are clearly seen radiating from the ridge in both simulations, in agreement with the properties of our configuration with a supercritical ridge and low excursion rate (see Section 2.2) and previous studies (Garrett and Kunze, 2007; Buijsman et al., 2010; Jalali et al., 2014). The 128-layer simulation has wave beams with more details and smaller structures when compared to the 8-layer simulation, which has much coarser vertical grid spacing; however, the ability of the 8-layer simulation to resolve the wave beams is still noteworthy. The presence of wave beams is due to the superposition of baroclinic modes (Gerkema, 2008).

It is notable that the peaks in baroclinic velocity, or internal wave beam bounces, both at the surface and at the bottom are not at the same place for the 8- and 128-layer simulations for the same time-snapshot – there is a lag in space. Further in the manuscript we show that this lag is due to differences among simulations in wavelength and phase speed for the first baroclinic mode. Values of wavelength and phase speed for the first baroclinic modes decrease with the increase in the number of layers up to 48 layers, and remain constant for further increases in the number of layers beyond 48. The spatial pattern of the time-averaged, depth-integrated baroclinic KE and APE across the domain is shown in Figure 4c,d. Due to the symmetry of the domain, only one half is shown and we provide a zoomed view within the first 250 km away from the center of the ridge.

The time-averaged, vertically-integrated baroclinic KE shows the same spatial pattern across simulations with different vertical resolutions: a peak at ~24 *km* away from the ridge, and a smooth decay further from it (Figure 4c). Roughly the same values of KE are found for all simulations independent of the grid-discretization and number of layers, except for the two simulations with 8 layers and, to some extent, the 16-layers simulations, at locations between 12 *km* to ~60 *km* away from the ridge. In contrast, APE magnitude differs more among simulations with different numbers of layers up to 48 layers, especially within the first 50 km from the ridge, and maintains similar magnitude in the simulations with higher layer counts (Figure 4d). With the addition of more layers beyond 48 layers, the results converge independent of the grid-discretization, highlighting the importance of adding isopycnal layers over modifying the grid-discretization. One maximum peak in APE is present at ~15 *km* with a value of 250 $J\ m^{-3}$ for simulations with at least 48 layers. For the two 16-layer simulations, the peak of maximum APE is 200 $J\ m^{-3}$ whereas the peak is as low as 160 $J\ m^{-3}$ in the 8 layers defined by the zero-crossings of the *u*-eigenfunction.

The baroclinic KE and APE peaks are not too far from the point of maximum steepness of the ridge (~20 *km*) and, consequently, the location of maximum tidal barotropic-to-baroclinic energy conversion. It is important to note that KE and APE are of the same order of magnitude and that both decay away from the ridge. The decay in tidal energy away from the source has been previously documented and attributed, among others, to nonlinear and wave-wave interactions (e.g., St. Laurent and Garrett, 2002; Lamb, 2004; Vic et al., 2019; Eden et al., 2020; Solano et al., 2023). Other potential causes of energy decay in our simulations could be attributed to linear wave dispersion and numerical mixing.

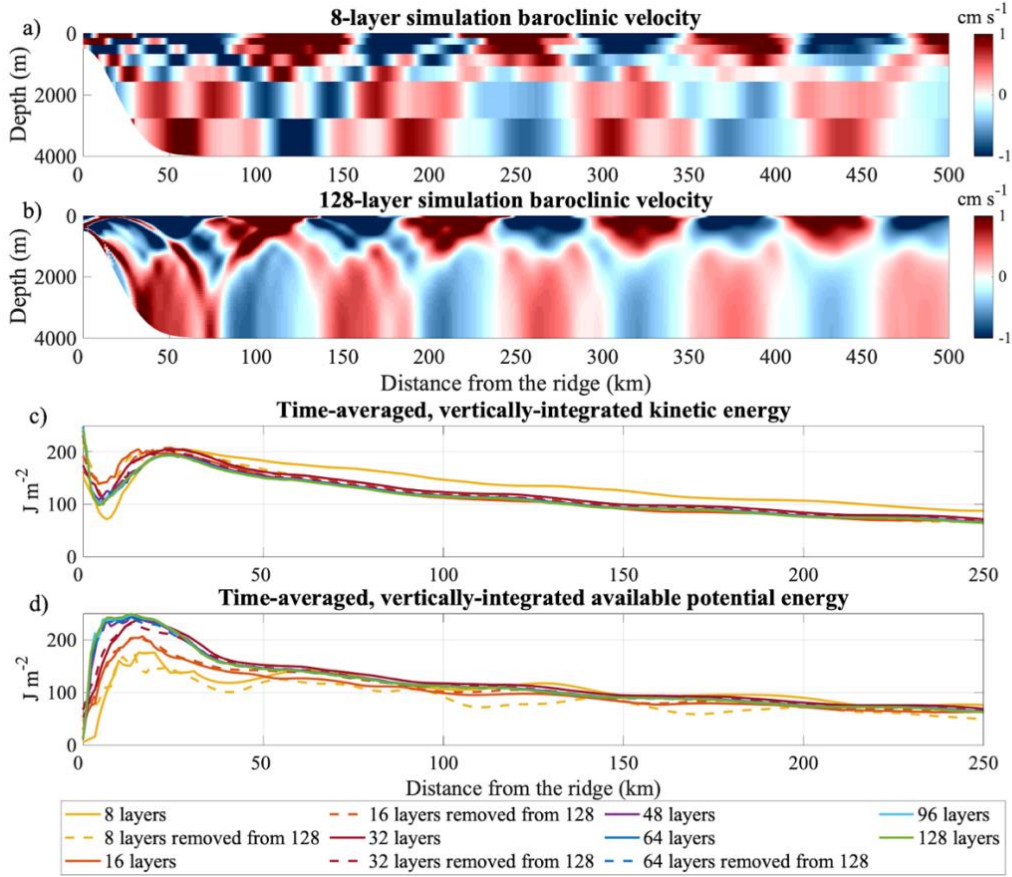

Figure 4. Snapshot of tidal baroclinic velocity for the a) 8-layer and b) 128-layer simulations forced solely by semidiurnal frequency, and with a 1/100° (~1 *km*) horizontal grid-spacing. Time-averaged, vertically-integrated (c) kinetic energy and (d) available potential energy. Solid lines are the simulations with layers defined using the zero-crossings of *u*-eigenfunctions, and the dashed lines are the simulations with layers defined by merging layers from the 128-layer simulation. Note that the x-axis range in c) and d) focuses on the first 250 *km* from the center of the ridge.

**3.2. Domain-integrated tidal energy conversion, baroclinic kinetic energy, and available potential energy**

Tidal barotropic-to-baroclinic energy conversion, APE, baroclinic KE, and vertical KE, integrated from the center of the ridge (x = 0) to 250 km from the ridge, for the two different vertical grid-discretization and different numbers of layers are shown in Figure 5. Tidal energy conversion differs slightly between simulations, with a small increase in averaged conversion with the increase in the number of layers for both sets of simulations until 32 layers, and it remains constant with further increase in layers (Figure 5a). Domain-integrated baroclinic kinetic energy, which is highly dominated by horizontal baroclinic kinetic energy by two orders or magnitude, decreases with the increase in the number of layers, independently of the grid-discretization (Figure 5b). The values go from $3.6 \times 10^{10}$ J for the 8-layer simulation defined by the zero-crossings of *u*-eigenfunctions, to $2.5 \times 10^{10}$ J for the 128-layer simulation. Note that the value of KE for the 8-layer defined by the zero-crossings of the 8th mode velocity eigenfunction stand out from the other simulations. We will see later that, for this simulation, less energy is being dissipated. The domain-integrated barotropic kinetic energy (not shown) is very similar among simulations with a slight decrease with the increase in the number of layers, from $1.0\times10^{10}$ J for the 8-layer simulations to $0.9\times10^{10}$ J for the 128-layer simulation.

The domain-integrated APE shows an increase in value with an increase in vertical layers until 48 layers, maintaining consistency in the simulations with higher layer counts ($3.2\times10^{10}$ J) independently of the grid-discretization (Figure 5c). This result agrees with findings from Figure 5b and highlights the importance of increasing the number of layers over adjusting the grid-discretization. Consistent with Figure 4, domain-integrated APE is of the same order of magnitude as domain-integrated baroclinic kinetic energy. Last, the domain-integrated vertical KE appears to increase with the increase in the number of layers in simulations having up to 32 layers, then slightly decrease until reaching constant vertical KE among simulations for those with at least 48 layers. While the vertical kinetic energy (½ $w^2$) is considerably smaller by two orders of magnitude than the horizontal baroclinic kinetic energy, it remains connected to the vertical displacement of isopycnals, which can impact underwater acoustic propagation (section 5) and mixing of water properties. This underscores the significance of vertical kinetic energy, prompting us to illustrate how this characteristic evolves with alterations in the number of layers.

Dissipation was estimated as the residual between the tidal barotropic-to-baroclinic energy conversion and the pressure force divergence (Table 1). Dissipation exhibits a similar pattern to the barotropic-to-baroclinic tidal convergence (Figure 5a); dissipation increases with the increase in the number of layers until 32-layers, then slight decreases for simulations with more than 48 layers, and remains constant after 96 layers. Notice the low value of dissipation for the 8-layer simulation defined by the zero-crossings of the $u$-eigenfunction compared to the dissipation of other simulations – this is in agreement with the higher values of baroclinic kinetic energy for this simulation (Figure 4 and Figure 5b).

Table 1. Dissipation (MW) estimated as the residual between the tidal barotropic-to-baroclinic energy conversion and baroclinic energy flux divergence integrated from 0 to 250 *km* from the center of the ridge. The vertical grid of the second set of simulations (second line in this table) is defined by merging subsequent layers from the 128-layer simulation defined by the zero-crossings of the $u$-eigenfunctions. Because of that, there is only one 128-layer simulation. The number of subsequent layers merged is a multiple of two (2, 4, 8, and 16), and gives rise to the second set of simulations with 64, 32, 16, and 8 layers, respectively. No simulations with 48 and 96 layers are present in this method since those numbers are not a multiple of 128.

| Method \ Number of layers | 8 | 16 | 32 | 48 | 64 | 96 | 128 |
|---|---|---|---|---|---|---|---|
| **Zero-crossings of u-eigenfunctions** | 1.0 | 2.6 | 3.4 | 3.5 | 3.4 | 3.3 | 3.2 |
| **Merging layers** | 2.1 | 2.9 | 3.5 | – | 3.4 | – | |

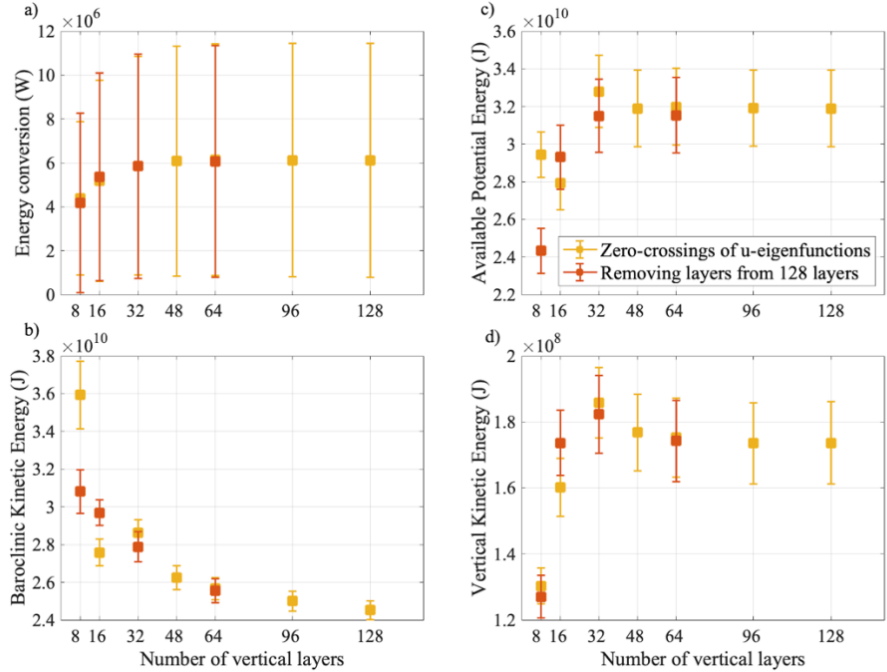

Figure 5. a) Domain-integrated tidal barotropic-to-baroclinic energy conversion ($W$), b) baroclinic (horizontal + vertical) kinetic energy ($J$), c) available potential energy ($J$), and d) vertical kinetic energy ($J$) as a function of the number of vertical layers for two grid-discretizations: layers defined using the zero-crossings of $u$-eigenfunctions (yellow), and layers defined by merging layers from the 128-layer simulation (orange). All the above were integrated spatially from 0 to 250 $km$ from the center of the ridge, and the standard deviation represents the temporal variability over four tidal cycles.

To understand the impact of numerical dissipation on the tidal energy, two model parameters were tested: the coefficient of quadratic bottom friction ($c_b$ – parameter name in HYCOM) and the diffusion velocity for biharmonic thickness diffusion (*thkdf4*). The first, as the name indicates, dissipates energy in the bottom mixed layer through bottom friction, and the second dissipates energy by smoothing isopycnal interfaces. In the simulations presented so far, $c_b$ and *thkdf4* are set as $2.5 \times 10^{-3}$ and 0.01 m s$^{-1}$, respectively, which are the standard for realistic and idealized HYCOM simulations. To test the influence of those parameters, twin simulations with 32 and 96 layers with isopycnal layers defined using the zero-crossings of $u$-eigenfunction were performed with different $c_b$ and *thkdf4*. The first set of experiments sets $c_b$ as zero while keeping *thkdf4* as 0.01 m s$^{-1}$, and in the second set of simulations $c_b$ is kept as $2.5 \times 10^{-1}$ and *thkdf4* is set to zero. When the bottom drag was set to zero, we found that the baroclinic kinetic energy and the APE for both 32- and 96-layer simulations increased by less than $0.01 \times 10^{10}$ J ($< 0.3\%$ of increase) [not shown]. When the biharmonic thickness diffusion was set to zero, the baroclinic KE and APE increased by less than $0.05 \times 10^{10}$ J ($< 1.5\%$ of increase) for both 32- and 96-layer simulations. The increases in baroclinic KE and APE associated with these two dissipation parameters were substantially smaller than the energy variation due to the number of layers.

### 3.3. Kinetic energy frequency spectra

The kinetic energy frequency spectra of the simulations with different numbers of vertical layers presented similar overall patterns, independent of the grid-discretization (Figure 6). It is important to note that even the simulations with only 8 layers reproduce all the peaks, which would be likely a challenge for models with only 8 z-levels, as highlighted by Buijsman et al., (2020, 2024) and Xu et al. (2023). The predominant internal tide frequency is semidiurnal (energy source). Wave-wave interactions lead to a transfer of energy to peaks at the frequencies multiple of the source frequency; in this case, peaks at $M_4$, $M_6$, $M_8$, $M_{10}$, and $M_{12}$ (e.g., Sutherland and Dhaliwal, 2022). Similar amplitude is seen for the $M_2$ frequency independent of the number of layers. At higher frequencies, nevertheless, less energy is found in simulations with a higher layer count.

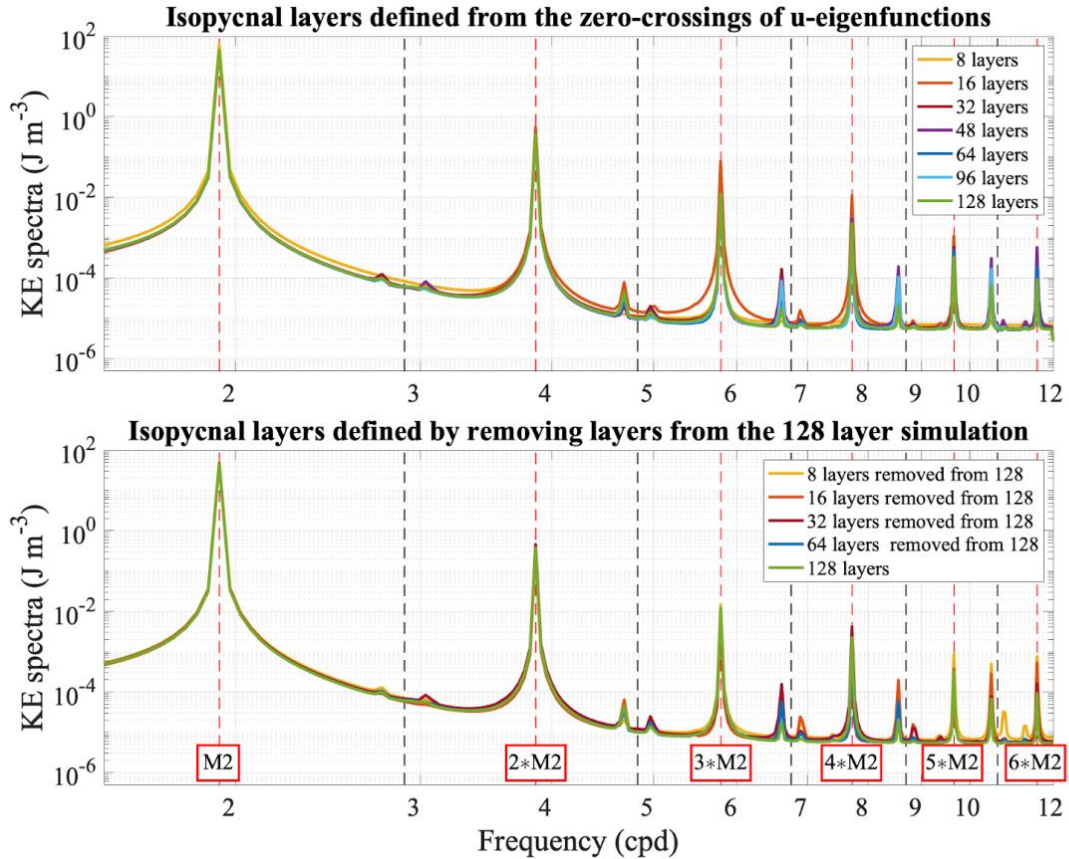

Figure 6. Baroclinic kinetic energy frequency spectra at 100 m depth, averaged over 0 to 250 km from the ridge, for different number of layers for grids defined (upper) by the zero-crossings of the *u*-eigenfunctions, and (lower) by merging intermediate layers from the 128-layer simulation. Note that the 128 layer simulation (green line) is the same in both sub-figures.

Peaks on each side of odd numbers of cycles per day ($M_3$, $M_5$, etc.) are observed in Figure 6, and believed to be associated with parametric subharmonic instability (PSI), also called Triad resonant instability (Varma and Mathur, 2017; Bourget et al., 2013). When PSI occurs, energy is transferred from a given input frequency to half that frequency, thus from higher to lower frequency. In domains with low viscosity, one peak at the half-frequency is observed; however, at higher viscosities, two peaks are observed, one on each side of the half-frequency. These peaks then interact with other frequency motions ($M_2$, $M_4$, $M_6$) and give rise to peaks at higher frequencies such as $M_3$, $M_5$, $M_7$, etc. PSI motions have very small horizontal scales in the order of *m* but low frequency. The two peaks on each side of the half-frequency are well documented and were observed in idealized simulations (Koudella and Staquet, 2006; Sutherland and Jefferson, 2020) and laboratory experiments (see Figure 2 in Joubaud et al., 2012; and Figure 3 in Bourget et al., 2013).

### 3.4. Squared vertical shear

The domain-integrated squared vertical shear increases with the increase in the number of layers for the two different vertical-grid discretization and levels off from the 96 to 128 layer simulations (Figure 7a). The difference is significant, with values ranging from ~ 425 $m^3s^{-2}$ for the 8-layer simulation defined using the zero-crossings of $u$-eigenfunctions to ~ 4100 $m^3s^{-2}$ for the 96- and 128-layers simulations. The 96- and 128-layer simulations presented high and sharp values of squared vertical shear in particular where wave beams are located (Figure 7b), contrasting from the lower resolution simulations that presented more diffuse and larger wave beams, leading to lower values of squared vertical shear (not shown). We notice lower shear values in the 32- and 64-layer simulations defined from merging intermediate layers compared to their counterparts defined from the zero-crossings of $u$-eigenfunctions. This discrepancy is likely related to the fact that the latter has more layers in the upper 200 $m$, better resolving the high-amplitudes of vertical shear in the upper-ocean.

High values of vertical shear associated with internal tides can lead to wave breaking, and subsequent mixing in the ocean. In fact, a recent study by Thakur et al. (2022) showed that models resolving internal tides exhibited higher vertical shear, eliminating the need for the background component of the K-Profile Parameterization (KPP) vertical mixing scheme. The work presented here shows the importance of choosing the grid-spacing carefully to best represent internal tide dynamics and consequent impacts on vertical shear, and potential consequences on mixing.

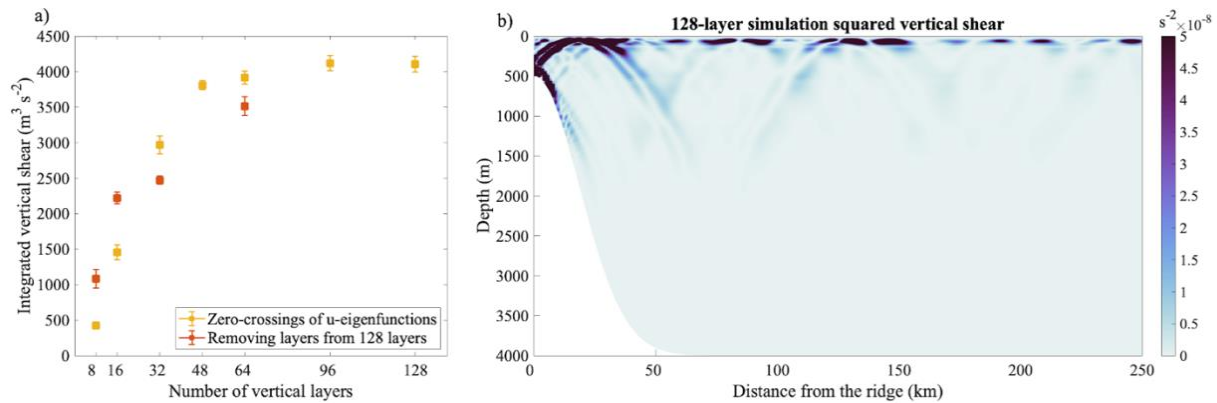

Figure 7. a) Same as 5a), but for the domain-integrated squared vertical shear. b) Snapshot of the squared vertical shear for the 128-layer simulation.

## 4. Modal kinetic energy and available potential energy

Modal KE and APE for baroclinic modes one to eight are shown in Figures 8 and 9. For all simulations, KE and APE projected only on the first eight vertical modes, even for the 8-layer simulations. This result agrees with the criteria proposed by Xu et al. (2023) and Buijsman et al. (2024), in which only $n$-layers are needed to resolve $n$ vertical modes. When comparing the total KE and APE with the sum of the modal KE and APE for the first 250 $km$ from the ridge, we find that less than 3% of the total KE and APE is not projected onto modes, with the exception of the 8-layer simulation defined by the merge of layers, for which 4.5% of the total KE was not projected onto the vertical modes. The residual (total KE and APE minus the sum of modal KE and APE) decreases with the increase in the number of layers.

For the first 100 *km* from the ridge, more than half (~60%) of the total modal KE is contained in the first baroclinic mode for all the simulations, ~7% is contained in the second baroclinic mode, and 17% in the third baroclinic mode (Figures 8a-c). From 100 km to 250 km from the ridge, around 80% of the total modal KE is contained in the first baroclinic mode for all simulations, ~3% in the second baroclinic mode, and ~11% in the third baroclinic mode. Modal KE converges for simulations with at least 32 layers independent of the grid-discretization. For the simulations with 8- and 16-layers, modal KE appears to be lower, on average, than the higher-resolution ones for the $3^{rd}$ and $5^{th}$ baroclinic modes, and larger for $4^{th}$ baroclinic mode. Most energy (> 90%) is contained in the first 5 modes, in agreement with Buijsman et al. (2024). The energy contained in higher modes decay rapidly away from the ridge, in agreement with previous studies (St. Laurent and Garrett, 2002; Lamb 2004; Vic et al., 2019; Eden et al., 2020; Solano et al., 2023).

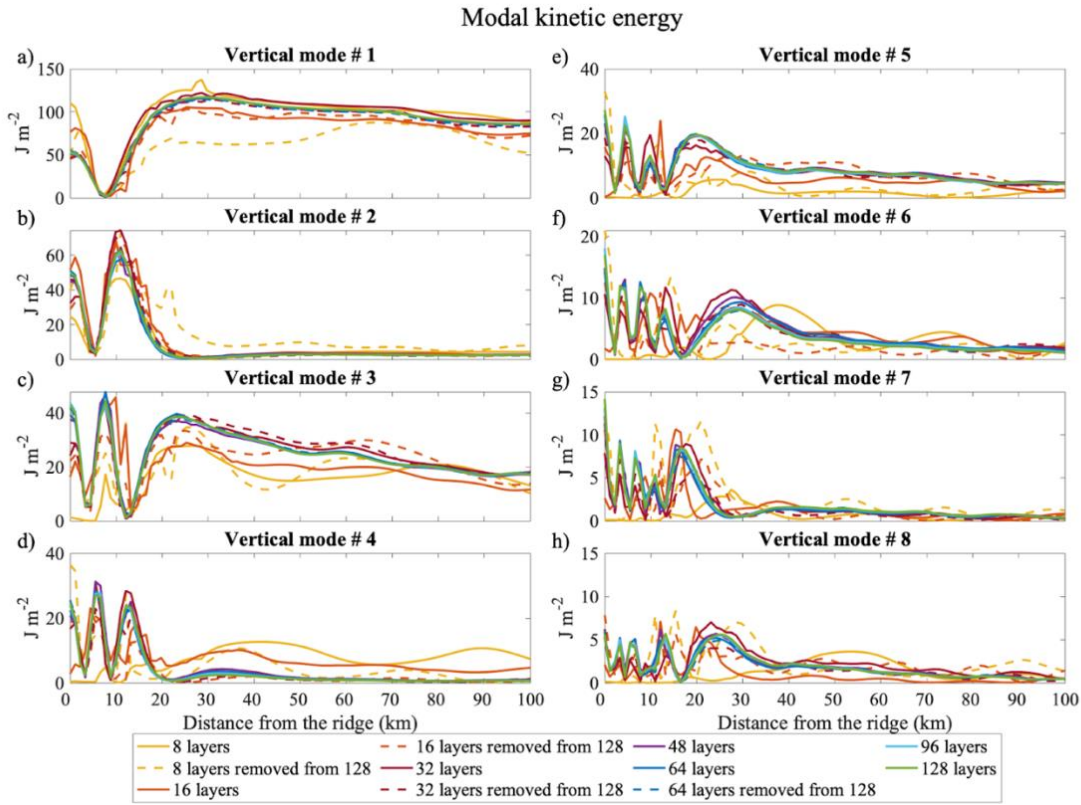

Figure 8. Vertically integrated, time-averaged contribution of each baroclinic mode on the total kinetic energy, from the 1st to the 8th baroclinic mode, for all simulations. Note that the ordinate axis varies for the different modes.

The differences caused by having different numbers of layers are more evident in the modal available potential energy (Figures 9). There is a clear increase in available potential energy with the increase in the number of layers, in particular for modes equal or higher than the 3rd baroclinic mode (Figure 9c-g). Modal APE converged for simulations with at least 48 layers independent of the grid-discretization. For the first 100 km from the ridge, and for simulations with at least 48 layers, 60% of the total modal APE is contained in the first baroclinic mode. This number increases with the decrease in the number of layers. On average, for the first 100 km from the ridge, around 7% is contained in the second baroclinic mode, and around 16% is contained in the third baroclinic mode (Figures 9a,c). From 100 km to 250 km from the ridge, around 84% of the total modal APE is contained in the first baroclinic mode for all simulations, ~3% in the second baroclinic mode, and ~10% in the third baroclinic mode (this value is smaller for the 8-layer simulations and the 16-layer one defined using the zero-crossings of u-eigenfunctions). The modal APE in the higher modes decayed rapidly away from the ridge compared to the energy associated with the first and third baroclinic modes. In summary, at least 48 vertical layers were needed to accurately resolve the APE associated with higher modes, in particular from the 3$^{rd}$ up to the 8$^{th}$ baroclinic mode. Both kinetic and available potential energy presented near zero values for modes higher than mode eight.

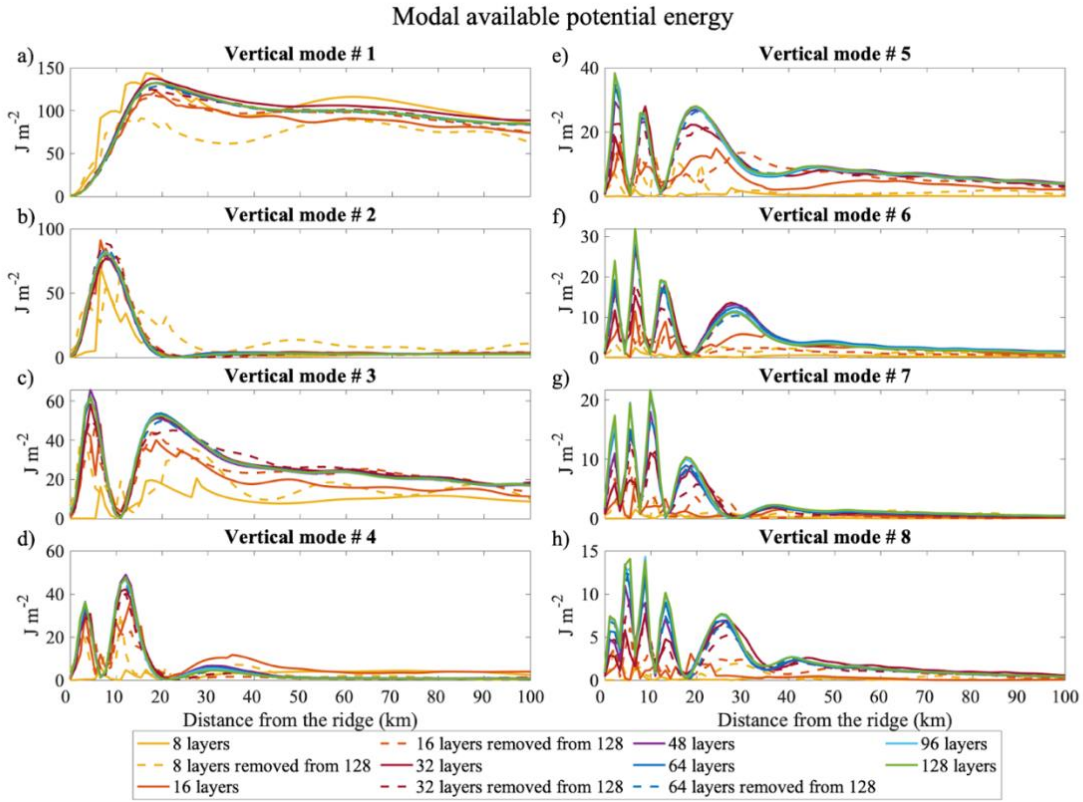

Figure 9. Same as Figure 9, but for modal available potential energy focused on the first 50 km from the center of the ridge.

Phase speed and wavelength of the baroclinic tides for modes one to ten, estimated using the Sturm-Liouville equation, are shown in Figure 10. Both phase speed and wavelength decreased with the increase in the number of layers until 48 layers, and remained constant with additional layers, independent of the grid-discretization. Phase speed and wavelength for each mode decreased from ~2.5 *m s$^{-1}$* and 110 *km* in the 8-layer simulations to 2.2 *m s$^{-1}$* and 100 *km* in the 48-layer simulation, respectively, for the first baroclinic mode. In fact, the wave beams were observed to be located closer to the ridge in simulations with at least 48 layers, and they extended further from the ridge when decreasing the number of layers (see Figure 4 for 8- and 128-layers), in agreement with Figure 10. The difference between phase speed and wavelength among simulations with different numbers of layers decreases with the increase in mode number.

The distance between subsequent wave beam bounces at the surface and bottom (~100 *km*; Figure 4) is similar to the wavelength of the first baroclinic mode (Figure 10). This means that internal wave beams bounce up and down over one mode-one wavelength for realistic stratification as well, similarly to the case of idealized, in-depth constant stratification in Gerkema and Zimmerman (2008).

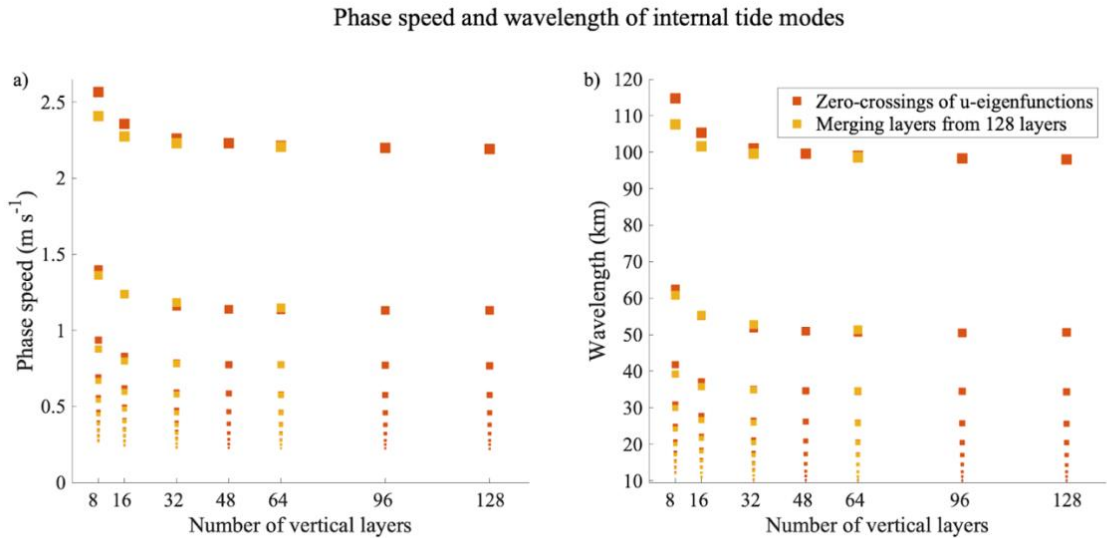

Figure 10. Internal tide a) phase speed and b) wavelength for the first ten baroclinic modes (from larger to smaller squares) for both types of grid-discretizations computed from the Sturm-Liouville equation.

## 5. Sound speed variability and underwater acoustics propagation

For this section, we focus on the simulations with layers defined by the zero-crossings of u-eigenfunctions, varying from 8 to 128 layers.

*5.1. Sound speed variability*

Sound speed depends on temperature, salinity, and depth. Thus, the up-and-down vertical displacement of isopycnals driven by internal tides induce sound speed variability. Findings from Section 3 show that simulations with a higher number of layers, up to 48 layers, presented higher vertical kinetic energy and available potential energy (Figure 5c, d). Both these quantities are also related to vertical isopycnal displacement. In fact, we find differences among the simulations in the vertical profiles of mean sound speed and an increase in sound speed variability with the increase in the number of layers, from 8 up to 48 layers, with the variability changing very little with additional layers (Figure 11). The 8-layer simulation presented a larger sonic layer, and, consequently, deeper sonic layer depth, and a deep channel (centered around ~1000 *m*) that is less pronounced compared to the other simulations with higher number of layers. With increase in layer count, the sonic layer depth decreases and the deep channel becomes more pronounced, with equilibrium after 48-layers.

The greatest variability, as measured by the standard deviation over each hourly time step for a total of four tidal periods, was found in the upper 250 *m* for all simulations, although the peak of maximum variability varied, with variability peaks at ~210 *m* depth for 8 layers and at ~110 *m* depth for 16 layers. For the higher-layer simulations (>16 layer), the peak sound speed variability was found at shallower depths (~60 m). Although the 32-layer simulation's depth of maximum variability was at ~60 *m*, the amplitude was 0.1 *m s$^{-1}$* larger compared to the simulations with higher layer count for 0-50 *km* range from the ridge. The larger amplitude in the 32 layers compared to the other simulations with higher layer count is persistent as we go further from the ridge (50-100 *km* and 100-250 *km*), but with a smaller difference in amplitude compared to 0-50 *km*. A secondary peak in sound speed variability was observed around ~1150 *m* with similar magnitudes for simulations with at least 32 layers. This peak is smaller in amplitude than the 16 layers and non-existent in the 8 layers. A similar pattern was found for the simulations with layers defined by merging consecutive layer layers from the 128-layer simulation defined by the zero-crossings of u-eigenfunction (not shown).

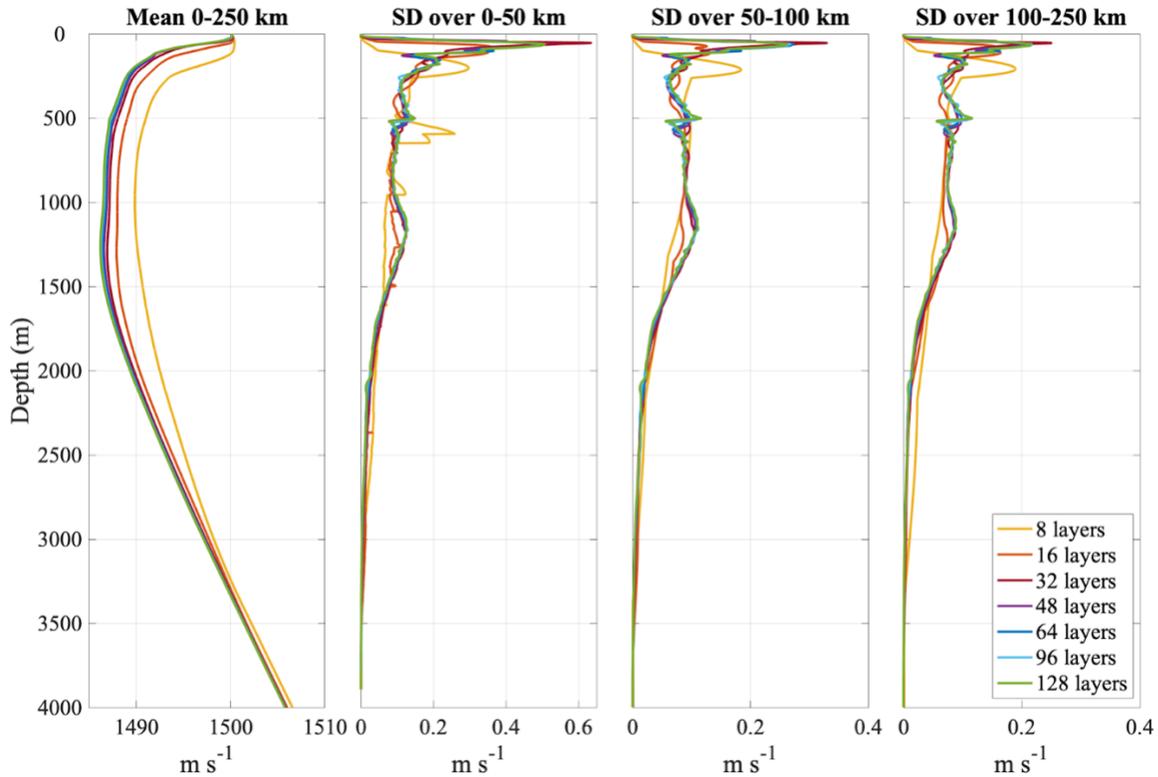

Figure 11. Mean sound speed averaged over 0-250 *km* from the center of the ridge, and sound speed standard deviation (SD) averaged over 0-50 *km* from the center of the ridge, 50-100 *km*, and 100-250 *km* for the simulations with 8, 16, 32, 48, 64, 96, and 128 layers defined by the zero-crossings of the *u*-eigenfunctions. Note that the range of the x-axis differs between the subplots.

*5.2. Underwater acoustic propagation*

The differences among simulations in sound speed average and variability shown in Figure 11 have an effect on the acoustic transmission loss (TL) and underwater acoustic parameters SLD, BLG, and ILG (Figure 12, 13; all terms are defined in Section 2.5). The transmission loss (1500 Hz source at 20 m depth at ridge location) was greatest for the 8-layer simulation, and decreased with the increase in the number of layers up to 64 layers, with no significant change for higher layer count (Figure 12). In the 64, 96, and 128 layer simulations, we observed a periodic transmission on a semidiurnal timescale, with two surface layer propagation pathways appearing at about every 12 hours. This signal is better explored in the next figure. For 48 layers, shorter periodicity was observed, whereas in the 32 layers, only one weak surface transmission was observed every ~12 hours. For 8 and 16 layers, surface layer propagation was weakest, with TL>75 dB.

The SLD becomes shallower with the increase in the number of layers up to 64 layers, reaching equilibrium afterwards (Figure 12). This result is in agreement with the sound speed averaged profiles from Figure 11a. The depth of the sonic layer varies from over 40 *m* for the 8 layer model and ~ 30 *m* for the 16-layer simulation, to ~ 20 *m* for the 32- and 48-layer simulations, and ~15 *m* for the simulations with at least 64 layers, in agreement with the mean profile of sound speed (Figure 11). For the BLG, simulations with at least 48 layers presented similar values of BLG, and larger gradient values compared to 8-, 16-, and 32-layer simulations. For ILG, the convergence happens at 64 layers, similarly to TL and SLD, with higher values for the simulations with at least 64 layers. Although at times the shallowness of the SLD for the higher-layer simulations means the source is deeper than the SLD, the gradients were more conducive to supporting the acoustic surface duct.

To expand on the acoustic transmission loss seen in Figure 13, we look at a depth-dependent snapshot of the transmission loss for all simulations (Figure 13). We observed a few differences among simulations. First, TL was greater for simulations with lower numbers of layers. As the number of layers increases, TL gradually diminished, reaching its lowest point in the 64-layer simulation. Beyond this point, TL stabilized, indicating a consistent behavior with the addition of extra layers, and in agreement with Figure 12. Second, the rays were more defined, sharper, and with higher density of rays/beams with the increase in the number of layers, up to 64 layers. The 8-layer simulation, for example, presents fewer propagation pathways, with little transmission in a deeper sound channel. As the number of layers increased, so did the number of propagation pathways. This caused greater propagation that converged in the deeper sound speed channel (centered ~1000 *m* depth), and nearer to the surface for higher numbers of layers, up to 64. Larger reflections (surface trapped sound waves) are observed in simulations with at least 64 layers. These reflections are nearly absent in the 8- and 16-layer ones for distances larger than 40 *km* from the ridge, and weaker in the 32- and 48-layer simulations. At a higher number of layers, results were similar.

The wavelength of the sound waves, easier seen by the distance between the surface and bottom wave bounces, appeared to be larger in the 8-layer simulation and to decrease with the increase in the number of layers up to 48 layers, similarly to the wavelength of the first baroclinic mode (Figure 10). Simulations with lower numbers of layers, in particular the 8- and 16-layer simulations, have a weaker deep sound channel (Figure 11a), which could be affecting the reflection angle of the sound waves and, consequently, their wavelength.

## 6. Conclusions

This study investigates the influence of vertical resolution on internal tide energetics and internal tide effects on underwater acoustics propagation in HYCOM. Two grid-discretizations with seven different numbers of layers, ranging from 8 to 128 isopycnal layers, were tested using a 2-D, idealized configuration, only forced with semidiurnal tides, with ~1-*km* horizontal grid-spacing and a ridge in the center of the domain. The results show that increasing the number of layers up to 48 layers increased the domain-integrated barotropic-to-baroclinic tidal conversion, available potential energy, and vertical kinetic energy, maintaining the same values in simulations with higher layer counts, independently of the grid-discretization. Domain-integrated vertical shear exhibits a similar pattern but reaching a maximum at 96 layers instead of 48, and remains constant with the addition of more layers. Domain-integrated kinetic energy, on the other hand, decreased with the increase in the number of layers. Simulations with at least 48 layers fully resolved the available potential energy contained in the $3^{rd}$ to $8^{th}$ tidal baroclinic mode. The wavelength and phase speed for the first ten baroclinic modes decreased with the increase in the number of layers, up to 48 layers. Thus, increasing the number of layers increased the amount of vertical structure in the flow, shown in the increase of energy in higher modes and squared vertical shear, both impacting internal tide-induced vertical transport.

Although differences were observed among simulations with lower number of layers, it is important to note that even the 8-layer simulations were able to reproduce internal tide patterns, such as wave beams and the 12 peaks in the kinetic energy spectrum. In the case of a z-level model, it would be needed at least three times that to obtain similar results as the isopycnal coordinates, according to Buijsman et al., (2020, 2024) and Xu et al. (2023). Additionally, the 8-layer simulations were able to resolve the same number of vertical modes (8 vertical modes) as the other simulations, even if the amount of energy projected onto the higher modes were smaller. This result agrees with the criteria proposed by Xu et al. (2023) and Buijsman et al. (2024), in which only *n*-layers are needed to resolve *n* vertical modes. This study shows that increasing the number of isopycnal layers is more efficient in terms of representing internal tide energetics over grid-discretization adjustments.

The increase in the number of layers also impact sound speed variability and underwater acoustic propagation. Acoustic analyses show an increase in sound speed variability, in particular in the upper 200 m, and subsequent changes in underwater acoustic properties, with addition of layers until 48 layers for sound speed variability, and 64 layers for underwater acoustic propagation, with not much changes observed with additional layers. We find that for simulations with less than 64 layers, the transmission loss was greater, and with less defined and more diffuse sound wave beams. The SLD and ILG both become shallower with the increase in the number of layers up to 64 layers. The same happens for BLG, but with a convergence at 48 layers instead of 64 layers.

Therefore, the study concludes that a minimum vertical resolution (roughly 48 layers in this case) is required to minimize the impact on internal tide energetics and internal-tide induced vertical transport and shear, both important to the mixing of water masses, and the subsequent consequences to sound speed variability and underwater acoustic propagation for the configurations of these simulations (1-km horizontal grid-spacing).

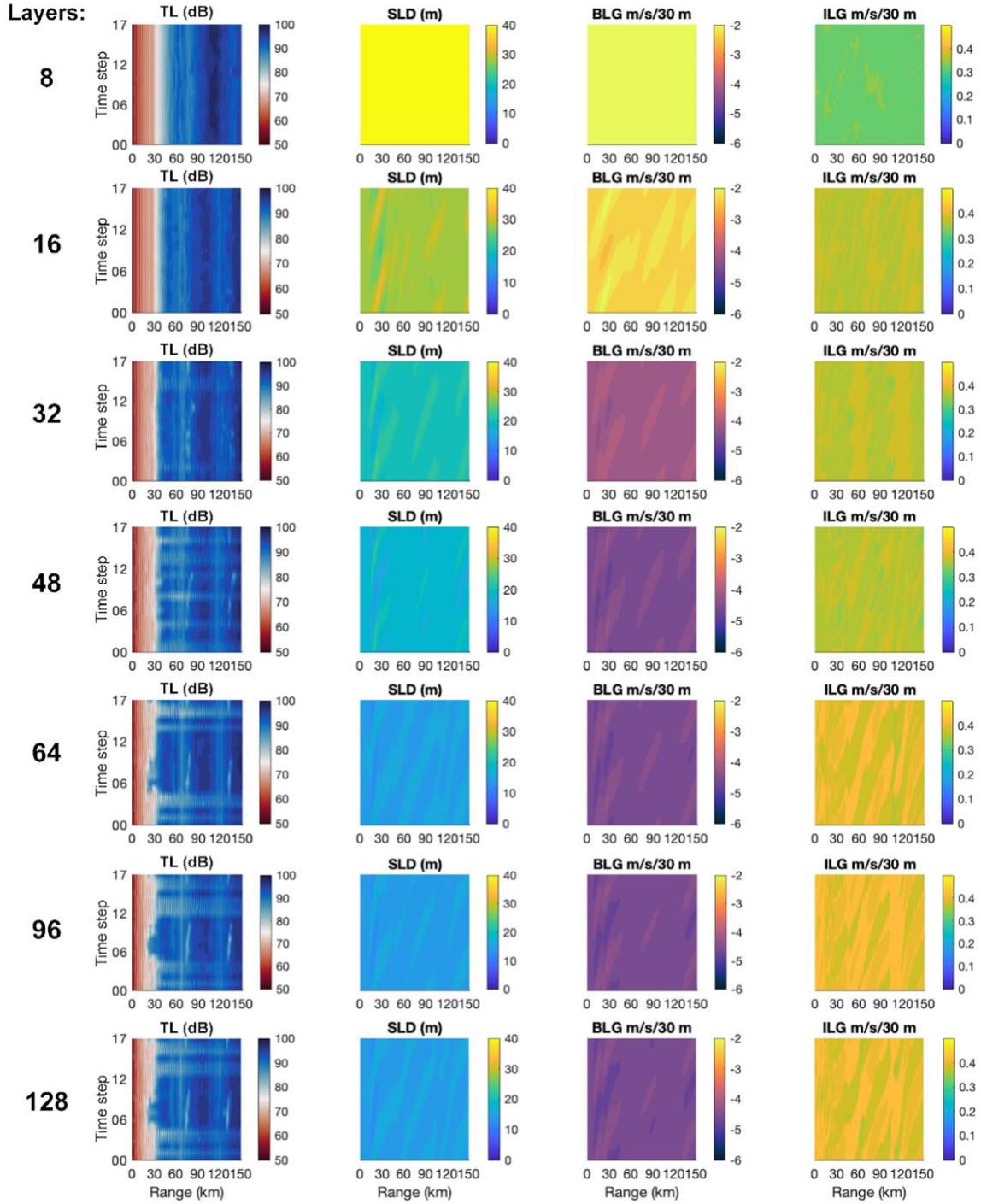

Figure 12. Underwater acoustic properties for the simulations with layers defined by the zero-crossings of *u*-eigenfunctions (from 8 to 128 layers, each line shows a different simulation). Acoustic transmission loss (TL; first column) in decibels (dB) with a 1500 Hz source at 20 m depth at ridge location; sonic layer depth (SLD; second column); below-layer gradient (BLG; third column); and in-layer gradient (ILG) defined as the gradient of sound speed in the sonic layer (fourth column). Sub-plots show the time step in the y-axis in hours and the distance from the ridge in kilometers (the zero value is the center of the ridge).

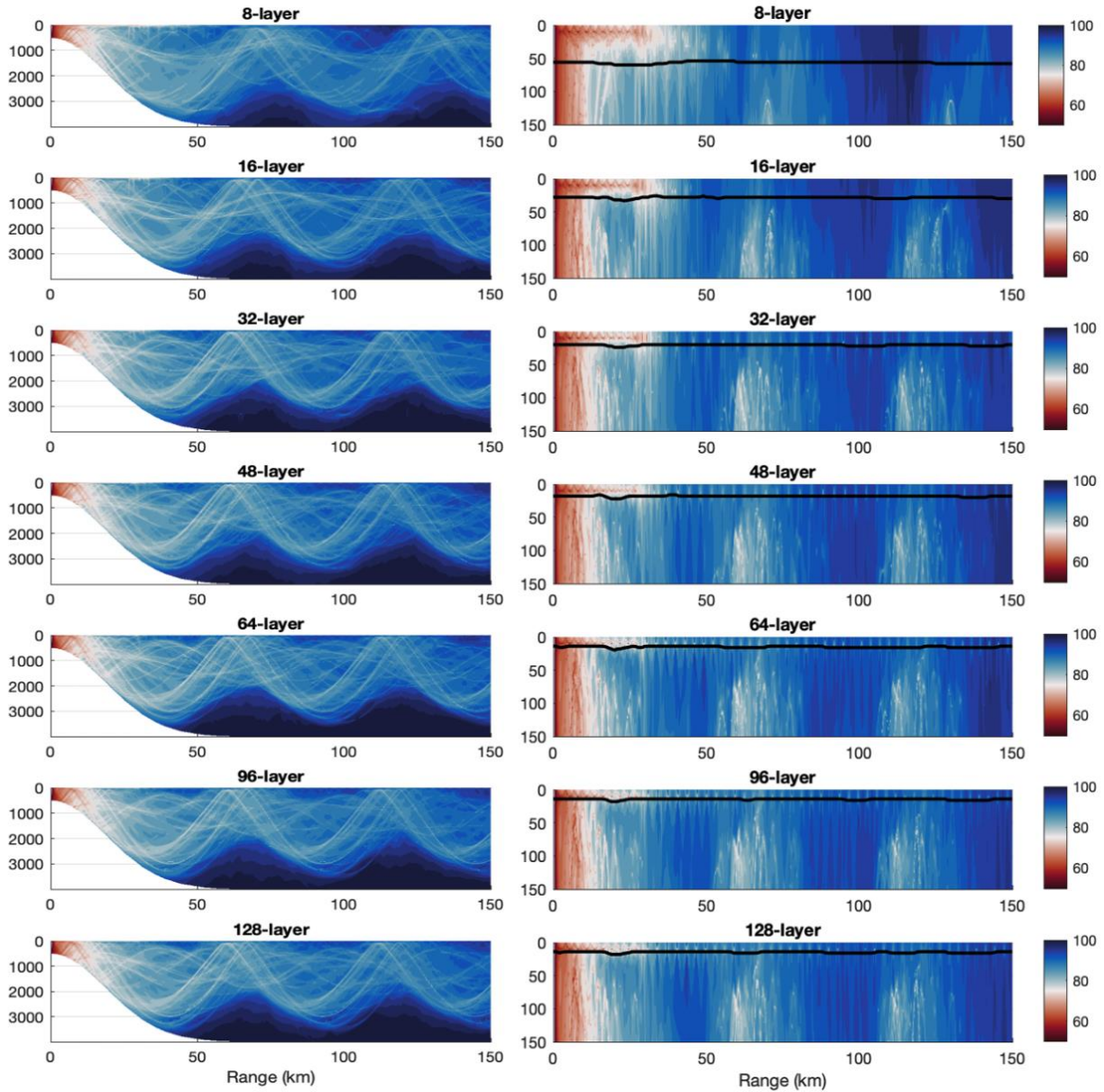

Figure 13. Snapshot of acoustic transmission loss (TL; first column) in decibels (dB) with a 1500 Hz source at 20 *m* depth at ridge location for the entire water column (left) and focused on the upper 150 *m* (right) for the simulations with layers defined by the zero-crossings of *u*-eigenfunctions (from 8 to 128 layers). The black line is the depth of the sonic layer.

## Acknowledgements

LH would like to thank Edward Zaron for his helpful comments, and Dheeraj Varma for the discussions on PSI. We acknowledge financial support from an Office of Naval Research (ONR) Task Force Ocean (TFO) project that the authors participated in. The ONR grant numbers for this TFO project are N00014-19-1-2717 (LH, KJR, EPC, AB), N00014-20-C-2018 (MS, EMCC),